\documentclass[trackchanges,twocolumn]{aastex701}
\usepackage{amsmath}
\usepackage{booktabs}
\usepackage{xcolor}

\begin{document}

\title{Discovery of sulfur oxides in the ejecta of a B[e] supergiant}

\author[orcid=0000-0002-7703-0692,sname='C. Bordiu']{Cristobal Bordiu}
\affiliation{INAF -- Osservatorio Astrofisico di Catania, Via Santa Sofia, 78, 95123 Catania (Italy)}
\affiliation{Instituto de Astrofísica de Andalucía (IAA-CSIC), Glorieta de la Astronomía s/n, Granada, Spain}
\email[show]{cristobal.bordiu@inaf.it}
\email[show]{cbordiu@iaa.csic.es}

\author[orcid=0000-0002-8443-6631, sname='Ricardo Rizzo']{J. Ricardo Rizzo} 
\affiliation{ISDEFE - Beatriz de Bobadilla 3, E-28040 Madrid, Spain}
\email[]{jrrizzo@isdefe.es}

\author[orcid=0000-0002-8499-7447]{David Navarro-Almaida}
\affiliation{Centro de Astrobiología, CSIC-INTA, Ctra. de Torrejón a Ajalvir km 4, 28850, Torrejón de Ardoz, Madrid, Spain}
\email[]{dnavarro@cab.inta-csic.es}

\author[orcid=0000-0001-6317-6343]{Asunción Fuente}
\affiliation{Centro de Astrobiología, CSIC-INTA, Ctra. de Torrejón a Ajalvir km 4, 28850, Torrejón de Ardoz, Madrid, Spain}
\email[]{afuente@cab.inta-csic.es}

\author[orcid=0000-0002-3429-2481]{Filomena Bufano}
\affiliation{INAF -- Osservatorio Astrofisico di Catania, Via Santa Sofia, 78, 95123 Catania (Italy)}
\email[]{filomena.bufano@inaf.it}

\author[orcid=0000-0002-6972-8388]{Grazia Umana}
\affiliation{INAF -- Osservatorio Astrofisico di Catania, Via Santa Sofia, 78, 95123 Catania (Italy)}
\email[]{grazia.umana@inaf.it}

\author[orcid=0000-0001-5126-1719]{Sara Loru}
\affiliation{INAF -- Osservatorio Astrofisico di Catania, Via Santa Sofia, 78, 95123 Catania (Italy)}
\email[]{sara.loru@inaf.it}

\author[orcid=0000-0002-1556-2474]{Alan C. Ruggeri}
\affiliation{INAF -- Osservatorio Astrofisico di Catania, Via Santa Sofia, 78, 95123 Catania (Italy)}
\email[]{alan.ruggeri@inaf.it}

\author[orcid=0000-0002-7288-4613]{Carla Buemi}
\affiliation{INAF -- Osservatorio Astrofisico di Catania, Via Santa Sofia, 78, 95123 Catania (Italy)}
\email[]{carla.buemi@inaf.it}

\author[orcid=0000-0003-1856-6806]{Francesco Cavallaro}
\affiliation{INAF -- Osservatorio Astrofisico di Catania, Via Santa Sofia, 78, 95123 Catania (Italy)}
\email[]{francesco.cavallaro@inaf.it}

\author[orcid=0000-0002-5537-7134]{Luciano Cerrigone}
\affiliation{Joint ALMA Observatory, Alonso de Córdova 3107, Vitacura, 7630355, Santiago, (Chile)}
\email[]{luciano.cerrigone@alma.cl}

\author[orcid=0000-0002-3137-473X]{Adriano Ingallinera}
\affiliation{INAF -- Osservatorio Astrofisico di Catania, Via Santa Sofia, 78, 95123 Catania (Italy)}
\email[]{adriano.ingallinera@inaf.it}

\author[orcid=0000-0003-4864-2806]{Paolo Leto}
\affiliation{INAF -- Osservatorio Astrofisico di Catania, Via Santa Sofia, 78, 95123 Catania (Italy)}
\email[]{paolo.leto@inaf.it}

\author[orcid=0000-0001-6368-8330]{Simone Riggi}
\affiliation{INAF -- Osservatorio Astrofisico di Catania, Via Santa Sofia, 78, 95123 Catania (Italy)}
\email[]{simone.riggi@inaf.it}

\author[orcid=0000-0002-1216-7831]{Corrado Trigilio}
\affiliation{INAF -- Osservatorio Astrofisico di Catania, Via Santa Sofia, 78, 95123 Catania (Italy)}
\email[]{corrado.trigilio@inaf.it}

\begin{abstract}
%Massive stars are the main drivers of galactic chemodynamical evolution, yet the synthesis and survival of molecules in their extreme environments remain poorly understood. In this context, 
B[e] supergiants represent a rare class of luminous, evolved massive stars surrounded by dusty circumstellar disks. Since their intense UV fields were long thought to sterilize their surroundings, molecular detections beyond carbon monoxide have remained elusive, leaving their chemical reservoirs largely unexplored. Whether these environments can sustain a complex molecular chemistry is a fundamental question with significant astrochemical implications. Here we report the detection of chemically rich molecular gas surrounding the B[e] supergiant HD~87643, using ALMA observations. Our data reveal the presence of the sulfur oxides SO and SO$_2$ and other sulfur-bearing species, marking the first detection of these molecules in an early-type evolved massive star. We find a high fractional abundance of SO$_2$ relative to H$_2$, which our chemical modelling can reproduce in timescales as short as $\sim$10$^4$ yr in an oxygen-rich environment. These results indicate that the detected molecules trace a short-lived, rapidly evolving phase of out-of-equilibrium chemistry. Furthermore, we measure an anomalously low $^{32}$SO/$^{33}$SO, that we attribute to mass-independent fractionation driven by intense photochemistry. This mechanism mirrors processes proposed to explain the $^{33}$S excesses in the atmosphere of the Archaean Earth. Our findings suggest that B[e] supergiants could serve as unique laboratories for studying sulfur chemistry under extreme radiation conditions, opening potential avenues to investigate the fractionation processes that shaped the isotopic signatures found in the early geological record.
\end{abstract}



\section{Introduction}\label{sec1}

B[e] supergiants (hereafter sgB[e]) are a rare class of luminous evolved massive stars ($\log L_\star/$L$_\odot \gtrsim 4$) exhibiting the B[e] phenomenon, i.e., a B-type spectrum with prominent low-excitation forbidden emission lines and strong infrared excesses from hot circumstellar dust \citep{all76,mir07}. Thought to be in a short-lived transitional evolutionary phase, sgB[e] stars undergo high mass-loss rates (10$^{-6}$--10$^{-4}$ M$_\odot$ yr$^{-1}$, \citealt{fre98}) and develop complex circumstellar environments (CSEs) of atomic gas, molecular gas and dust, whose formation process remains little understood \citep{hil06}.

Their hybrid spectra---combining narrow and broad emission lines, P Cygni or double-peaked profiles, and intrinsic polarization---suggest multiple, well-differentiated emitting regions. To account for this phenomenology, \cite{zic85} proposed a two-component wind model for sgB[e] stars, consisting of a fast, line-driven polar wind and a slower, cooler equatorial outflow. With an estimated density contrast between components of $\sim$100--1000 \citep{zic89}, the denser equatorial outflow would form a disk-like structure where dust condenses, giving rise to both forbidden and molecular emission. This model has since become the leading paradigm to explain the non-spherical envelopes of sgB[e] stars, supported by substantial observational evidence, including polarimetric studies \citep{mag92,ser17}, detections of molecular emission, mostly K-band CO band heads \citep{mcg88,mar18}---with much less frequent SiO, \cite{kra15}---and even spatially resolved imaging of disks using near-infrared interferometry \citep{mil09, whe12}. However, despite these advances, the formation, composition, and kinematics of this circumstellar material remain unclear. In this context, the millimeter regime---still largely unexplored for this class---offers a unique window onto sgB[e] stars: it probes the cool dust and the chemistry of their equatorial outflows at distances beyond the hot innermost regions accessible in near-IR, placing firm constraints on their mass budget.

Among the limited census of Galactic sgB[e] stars \citep{kra19}, {HD~87643} (=V640 Car, Hen 3-365) is notable for its exceptionally strong infrared excess, one of the largest in its class ($J-K$ $\sim$ 2.6, \citealt{mac88}). Located at a distance of $d=1.6$ kpc\footnote{\textit{Gaia} DR3 parallax is $\epsilon=$0.6293$\pm$0.1227 mas, corresponding to a distance of 1.58$^{+0.39}_{-0.25}$ kpc. See Appendix \ref{methods:evol} for more details.}, it is a B3I star \citep{sho90}  with an estimated mass of $\sim$25 M$_\odot$ \citep{oud98}. {HD~87643}  is surrounded by an extended and highly asymmetric reflection nebula, bright in H$\alpha$ \citep{ber72}. The presence of P Cygni Balmer profiles suggests that the nebula is the result of significant mass-loss from the central star \citep{sur81}. Subsequent modeling of spectroscopic and polarimetric data by \cite{oud98} revealed a fast wind in excess of $\sim$1000 km s$^{-1}$ and a slowly rotating, expanding equatorial component traced by forbidden lines, consistent with the overall picture for sgB[e] stars proposed by \cite{zic85}. 

Years later, VLT/AMBER interferometric observations by \cite{mil09} resolved {HD~87643} as a binary system with a projected angular separation of 35$\pm$5 mas ($\sim56\pm8$ au at 1.6 kpc). The system consists of an early-type supergiant surrounded by an oxygen-rich circumprimary disk, and a fainter companion, likely with its own dusty envelope. The two stars appear embedded in a cool circumbinary dusty envelope, indirectly inferred to have a size of up to $\sim$1 arcsec ($\sim$1600 au). For decades, K-band CO emission  remained elusive in {HD~87643} \citep{mcg88}, until its unambiguous detection by \cite{mar18}, confirming the presence of hot molecular material in the circumprimary disk.

Here, we used the ALMA's Atacama Compact Array (ACA) to perform the first millimetric survey of HD~87643 and its circumstellar environment. We report the detection of chemically rich molecular gas surrounding the star, with CO, $^{13}$CO and multiple sulfur-bearing species, marking the first confirmed indication of sulfur chemistry within the sgB[e] class.

\section{Observations and data reduction}
\label{methods:observations}

HD~87643 was observed as part of the ALMA cycle 10 program ULISSES (Unbiased LIne Survey of Supergiant Evolved Stars, proj. 2023.1.01688.S, P.I: C. Bordiu). The observations were carried out using the 7-m array, employing all 11 antennas, across five nights in May 2024 (10th, 13th, 20th, 27th and 28th), for a total on-source integration time of 2.9 hours. Bandpass and flux calibration were performed using calibrators J1107-4448 and J1037-2934, while J0904-5735 served as the phase and pointing calibrator.

A 90"$\times$90" mosaic was obtained around the star coordinates ($\alpha=10^h04^m 30.28^s, \delta=-58^\circ 39' 52.09''$), to cover the full extent of the H$\alpha$ nebula around {HD~87643}. The spectral setup comprised four spectral windows centered at 218.404, 220.398,  230.538, and 231.901 GHz. The velocity resolution varied from 0.7 to 1.3 km s$^{-1}$\,.
Calibration and imaging of the raw visibilities were performed using CASA v6.5.4.9 and the ALMA pipeline v2023.1.0.124. The final data products include a continuum map and four spectral cubes, with a characteristic synthesized beam of 7.1"$\times$4.7" and a line rms better than 30 mJy beam$^{-1}$. Subsequent analysis was carried out with \texttt{CARTA} (v4.0.0) and \texttt{GILDAS}.

\begin{table}[h!]
\centering
\caption{List of transitions detected, with their respective rest frequencies and lower level energies.}
\label{tab:detections}
\begin{tabular}{llcc}
\hline
\hline
Species & Transition & $E_\mathrm{L}$ (K) & $\nu_0$ (GHz) \\ 
\hline
$^{33}$SO       & $6_5 \rightarrow 5_4, F=15/2 \rightarrow 13/2$ & 24.22  & 217.833 \\
H$_2$CO         & $3_{0,3} \rightarrow 2_{0,2}$                  & 10.48  & 218.222 \\
H$_2$CO         & $3_{2,2} \rightarrow 2_{2,1}$                  & 57.61  & 218.476 \\
H$_2$CO         & $3_{2,1} \rightarrow 2_{2,0}$                  & 57.61  & 218.760 \\
OCS             & $18 \rightarrow 17$                            & 89.30  & 218.903 \\
HNCO            & $10_{1,10}\rightarrow 9_{1,9}$                 & 90.57  & 218.981 \\
SO$_2$          & $22_{7,15}\rightarrow 23_{6,18}$               & 342.23 & 219.276 \\
SO $^{3}\Sigma$ & $6_5\rightarrow 5_4$                           & 24.43  & 219.949 \\
$^{13}$CO       & $2\rightarrow 1$                               & 5.29   & 220.398 \\
O$^{13}$CS      & $19\rightarrow 18$                             & 99.48  & 230.317 \\
$^{12}$CO       & $2\rightarrow 1$                               & 5.53   & 230.538 \\
OCS             & $19\rightarrow 18$                             & 99.81  & 231.061 \\
$^{13}$CS       & $5\rightarrow 4$                               & 22.19  & 231.221 \\
\hline
\end{tabular}
\end{table}

%Molecular line identification was performed using the \textit{Spectral Line Query} utility within \texttt{CARTA}, which provides access to the CDMS \citep{mul05} and JPL \citep{pic98} spectroscopic databases. We initially identified the $^{12}$CO and  $^{13}$CO $J=2\rightarrow1$ transitions since these lines appear isolated in the spectra. The velocity of the brightest component ($-18$ km s$^{-1}$) was then adopted as a kinematic reference for identifying the remaining lines---which consistently show a single velocity component. 

%$^{33}$SO  is the only species showing hyperfine structure, but the moderate velocity resolution of our data prevents us from completely resolving the individual hyperfine components. In cases where multiple candidate species were possible for a given frequency, assignments were prioritized based on the lower-level energies ($E_L$) and the relative intensities of the transitions, to ensure physical consistency with the environment of HD~87643. All reported detections exceed the $3\sigma$ threshold, with the majority of lines detected above 5$\sigma$.

\section{Results}\label{sec2}

%To put this detections in context, we employed radiative transfer modelling and incorporated archival photometry data to reconstruct the spectral energy distribution (SED) of HD~87643. This, coupled with chemical kinetics models, allowed us to paint a comprehensive physico-chemical portrait of the circumstellar environment of this source. Our analysis identified several chemical anomalies, and showed that the molecules were likely formed during the post-Main Sequence evolution of the star, possibly triggered by an episode of enhanced mass-loss.

\subsection{Overview of the detections}\label{subsec2}

An initial inspection of the data products clearly revealed unresolved continuum emission at the position of {HD~87643}, with a flux density $S=216\pm2$ mJy.  Coincident with the continuum, we also identified emission from ten different molecular species, namely CO, $^{13}$CO, OCS, H$_2$CO, SO, $^{33}$SO, SO$_2$, HNCO, O$^{13}$CS, $^{13}$CS, totalling thirteen spectral lines. The $^{12}$CO and  $^{13}$CO $J=2\rightarrow1$ transitions were adopted as a kinematic reference for identifying the remaining lines. Quantum numbers, lower-level energies, and rest frequencies for all detected transitions are presented in Table \ref{tab:detections}.

\begin{figure*}
    \centering
    \includegraphics[width=0.95\textwidth]{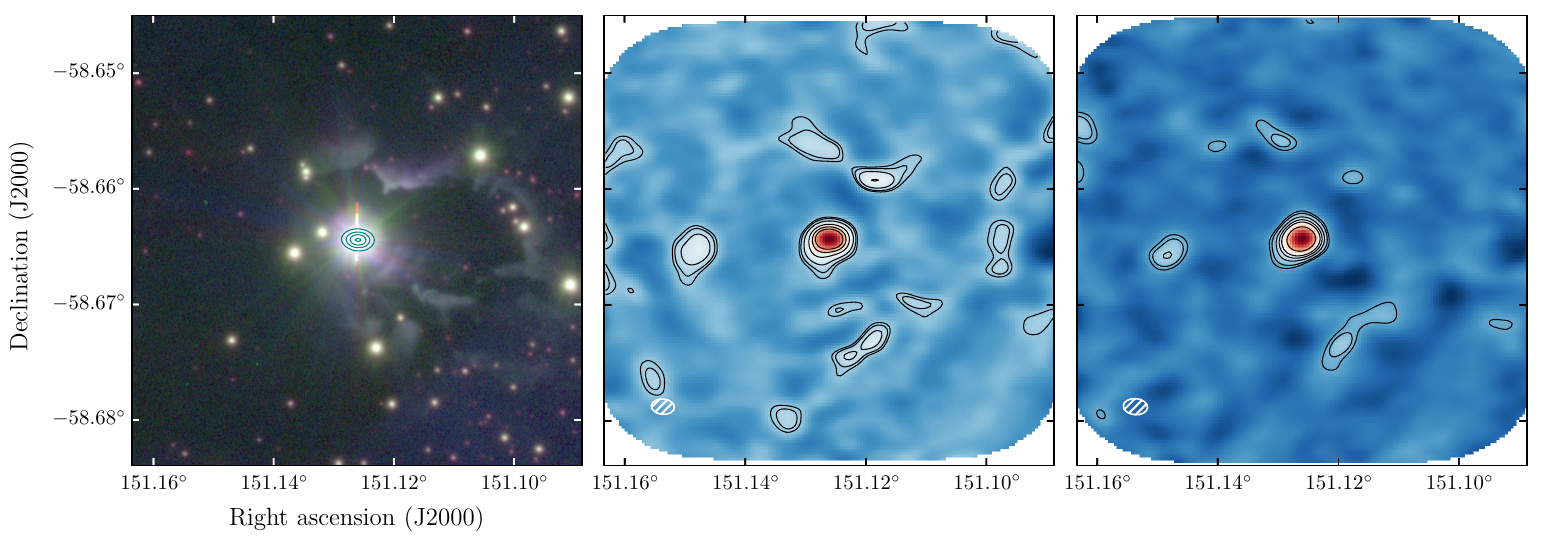}
    \caption{CO and $^{13}$CO emission in the field of {HD~87643}. \textit{Left}: RGB composite image from VPHAS+ showing the star and its surrounding nebula, with H$\alpha$ mapped to blue, $u+g$ to green, and $r+i$ to red. ALMA 1.3 mm continuum contours are overlaid in green at levels of 50, 100, 150, and 200 mJy beam$^{-1}$. \textit{Center}: CO $J=2\rightarrow1$ moment 0 map integrated over the velocity range $-32$ to $-5$ km s$^{-1}$, with contours at 15, 20, 30, 50, 70, and 90 Jy beam$^{-1}$ km s$^{-1}$. \textit{Right}: Same as center, for $^{13}$CO $J=2\rightarrow1$, with contours at 6, 8, 12, 16, 20, and 30 Jy beam$^{-1}$ km s$^{-1}$. The dashed ellipse in the bottom-left corner indicates the ALMA synthesized beam.}
    \label{fig:fig1}
\end{figure*}

\begin{figure*}
    \centering
    \includegraphics[width=0.95\textwidth]{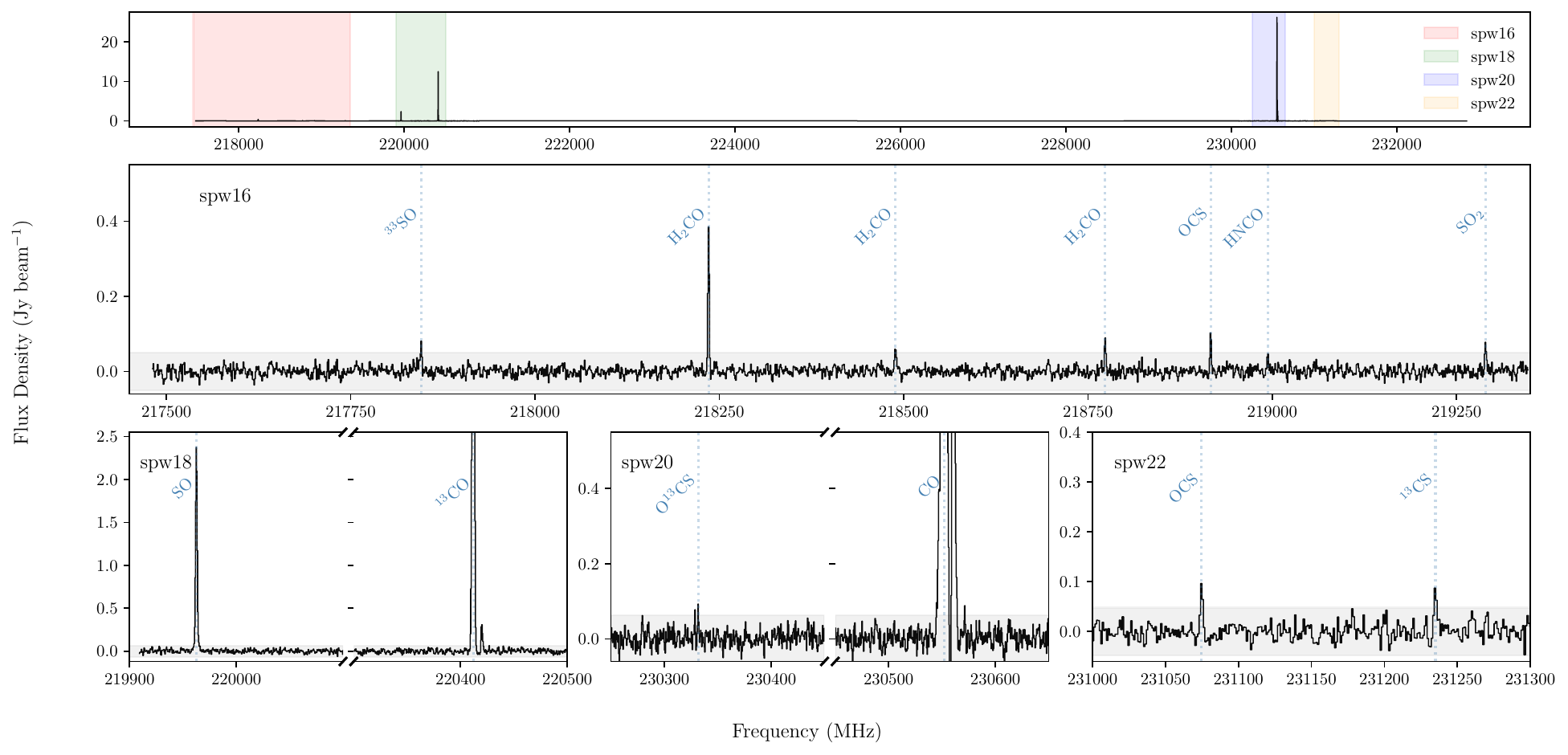}
    \caption{Beam-averaged spectra toward {HD~87643}. The top panel displays the full spectral coverage, with zoomed views of individual spectral windows below. The x-axes of spectral windows 18 and 20 are broken to omit line-free regions. Shaded grey areas indicate the 3$\sigma$ noise level. Blue dashed lines indicate identified molecular species at a velocity of $-18$ km s$^{-1}$. }
    \label{fig:fig2}
\end{figure*}

The spectra of CO and $^{13}$CO integrated over the entire field show most of the emission is concentrated within the velocity range $(-32, -5)$ km s$^{-1}$, consistent with the location of the star on the near side of the Carina arm. Fig. \ref{fig:fig1} presents the integrated emission maps of CO and $^{13}$CO over this  range, alongside a VPHAS+ RGB composite image of the star and the nebula, overlaid with the 1.3 mm continuum from ACA. The maps reveal two distinct structures that appear to be physically associated with {HD 87643}: 

\begin{enumerate}
    \item A series of dispersed clumps surrounding the star, spread over the entire velocity range and particularly prominent in CO. These clumps trace the brightest regions of the optical nebula, including the northern edge and the southern \lq\lq spur\rq\rq (labelled as B1 and B3 in \citealt{mil09}, fig. 6); the only non-coincident structure is a bright clump located $\sim$40 arcsec east of the star, clearly detected in both lines but lacking an optical counterpart, likely due to dust obscuration.
    \item A  central emission component originating at the position of the star, visible in both CO and $^{13}$CO. This component has a central velocity of $-18$ km s$^{-1}$, which is compatible with the systemic velocity of HD~87643 determined by \cite{oud98} of $-17\pm$4 km s$^{-1}$\, from the Fe\textsc{ii} and [O\textsc{i}] lines. This central emission is  notably brighter than the outer clumps and is slightly resolved by the ACA beam, extending beyond the millimeter continuum source. Fitting an ellipse to the half-maximum contour, we determine the structure has an angular size of $\sim7.5\times10$ arcsec, corresponding to $\sim12000\times16000$ au at the distance of {HD~87643}.
\end{enumerate}

In contrast to CO, all other detected molecular species appear exclusively at the stellar position. They are unresolved and show a single velocity component at $\sim-18$ km s$^{-1}$, coincident with that of CO and $^{13}$CO. This strong correspondence in position and velocity confirms their association with the central environment of {HD~87643}. Fig. \ref{fig:fig2} presents beam-averaged spectra for the four observed spectral windows at this position, indicating the detected transitions. 

These detections substantially expand the molecular inventory of {HD~87643}, previously limited to CO,  and more importantly, establish a new chemical benchmark for early-type supergiants: this work marks the first detection of sulfur-bearing species towards a sgB[e] star. Sulfur, with a cosmic elemental abundance S/H $\sim  1.5\times10^{-5}$, is the tenth most abundant element in the Universe and one of the six fundamental ingredients of life \citep{mif21}. While sulfur species have been found in a variety of astrophysical environments---from the Solar System and protoplanetary disks to star-forming regions and the envelopes of some evolved stars ---their formation and survival in the ejecta of hot, evolved massive stars is yet to be established; to date, only SiS has been reported in the environment of the luminous blue variable (LBV) star $\eta$ Car \citep{bor22}. Therefore, these new detections in {HD~87643} open new avenues to explore the sulfur chemistry in such unique environments, motivating the search for sulfur-bearing species in similar sources.

\subsection{Origin of the molecular emission}\label{subsubsec2}
 
Previous studies of  {HD~87643} revealed a complex circumstellar environment: a resolved circumprimary disk and an unresolved companion embedded in dense, cooler circumbinary material \citep{mil09}. Subsequent near-infrared observations detected CO $v=2\rightarrow0$ bandhead emission at  2.29 $\mu$m with a line-of-sight projected rotational velocity $v_\mathrm{rot} \sin{i} =11\pm1$ km s$^{-1}$ \citep{mar18}. To reconcile such a low velocity with the high temperatures required to excite the CO emission (between 2000 and 5000 K), the authors proposed that the molecular gas originated in the compact circumprimary disk  ($r\sim$3 au), necessarily observed at a low inclination angle ($i=7$ deg). While these near-IR observations traced the innermost, hottest molecular gas, our detections probe a much cooler gas component located farther away from the stars. Indeed, the CO emission is clearly resolved, implying that at least a fraction  of the gas must have a circumbinary origin (as the ACA beam samples spatial scales two orders of magnitude larger than the binary separation of $\sim$56 au). In this environment CO can survive the intense UV radiation thanks to a combination of self-shielding---where the innermost gas layers protect the material further out---and dust shielding within the circumbinary envelope posited by \cite{mil09}.  We thus conclude that the CO emission detected by ALMA is related to this larger circumbinary structure.

As for the other detected molecules, their origin is less certain, since we lack the spatial or velocity resolution to unambiguously determine their location and kinematics. However, any emission from the compact circumprimary disk would likely suffer from significant beam dilution, so the circumbinary origin remains the most plausible scenario for all the species. This is further supported by the comparable linewidth of all the lines, which suggests the emission arises from the same volume. We adopt this as our working hypothesis in the remainder of this work.

\subsection{Fractional abundances}

To derive first-order estimates of the molecular column densities, we used the  software package \texttt{MADCUBA} \citep{mar19}.\texttt{MADCUBA} determines the physical parameters of the molecular emission (column density $N$, excitation temperature $T_\mathrm{ex}$, linewidth FWHM, and central velocity $V_\mathrm{LSR}$) by fitting synthetic LTE models to the observed spectra. Specifically, the \textsc{AUTOFIT} function performs a nonlinear least-squares fit using the Levenberg-Marquardt algorithm \citep{lev44, mar63},  returning the parameters that best reproduce the line profiles.

When only a single transition is available, as for most detected species, the degeneracy between $T_\mathrm{ex}$ and $N$ prevents an unambiguous solution and makes it difficult to constrain line opacity, even though \texttt{MADCUBA} accounts for this effect. The exceptions are OCS and H$_2$CO, with two and three
detected transitions, respectively. Although the narrow range of upper energy levels is insufficient to tightly constrain $T_\mathrm{ex}$, they yield coarse estimates of 56 and 48 K, respectively.

\begin{figure*}
    \centering
    \includegraphics[width=0.75\textwidth]{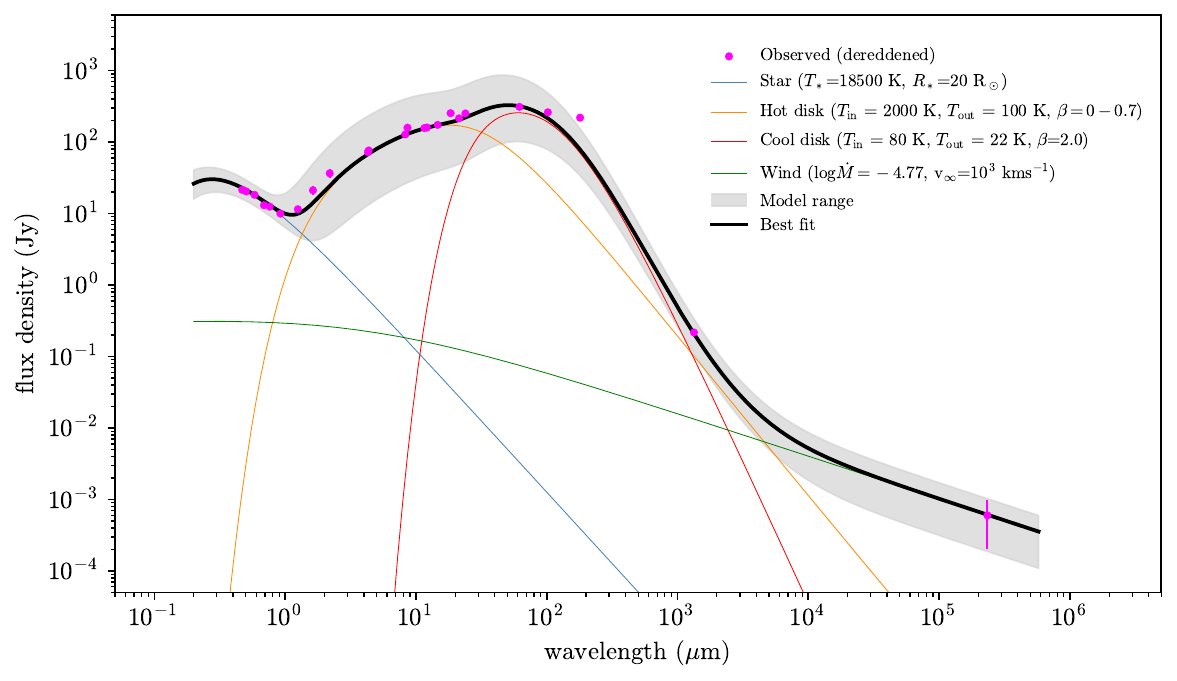}
    \caption{Spectral energy distribution of HD~87643. The magenta points show the dereddened photometric data with associated error bars. Individual components of the best-fit model are indicated by the colored lines: stellar blackbody (blue), hot and cold disks (orange and red), and stellar wind (green). The combined model is shown in black, with the uncertainty represented by the grey shaded region.}
    \label{fig:fig3}
\end{figure*}

We thus decided to fix $T_\mathrm{ex}$ a priori for all species. To make a more informed guess of the temperature, we modeled the spectral energy distribution (SED) of {HD 87643} from optical to radio wavelengths using literature photometry and our flux density measurements (see Appendix \ref{methods:sed} for details). The resulting fit, shown in Fig. \ref{fig:fig3}, successfully reproduces the observed emission and yields $T_\mathrm{dust}$=20--80 K for the circumbinary envelope, compatible with the OCS and H$_2$CO estimates. We therefore assumed LTE and full thermal coupling, such that $T_\mathrm{dust}=T_\mathrm{gas}=T_\mathrm{kin}=T_\mathrm{ex}$. Furthermore, lacking spatial information on the molecular distribution, we assumed that all molecular emission is co-spatial and fills the beam. Under these assumptions, we reproduced the molecular emission using two representative excitation temperatures, namely 50 and 100 K. The latter accounts for potentially warmer material
between the circumprimary and circumbinary structures. Temperatures below 30-40 K are unlikely, as they produce heavily saturated CO and $^{13}$CO line profiles inconsistent with the observations. The fitting results are presented in Table \ref{tab:lte}. 

\begin{table*}[]
\centering
\caption{\texttt{MADCUBA} line fitting parameters, column densities and fractional abundances for the detected molecules, assuming excitation temperatures of 50 and 100 K.}
\label{tab:lte}
\begin{tabular}{lcccccc}
\hline
\hline
Species & $V_\mathrm{LSR}$ & FWHM & $N_\mathrm{50K}$ & $X(N/N_\mathrm{H_2})_\mathrm{50K}$ & $N_\mathrm{100K}$ & $X(N/N_\mathrm{H_2})_\mathrm{100K}$ \\
 & (km s$^{-1}$) & (km s$^{-1}$) & (cm$^{-2}$) & & (cm$^{-2}$) & \\
\hline
CO          & $-18.1\pm0.1$ & $3.7\pm0.1$ & $(7.59\pm0.17)\times10^{16}$ & $10^{-4,a}$        & $(1.15\pm0.03)\times10^{17}$ & $10^{-4,a}$ \\
$^{13}$CO   & $-18.2\pm0.1$ & $2.9\pm0.1$ & $(2.57\pm0.79)\times10^{16}$ & $3.3\times10^{-5}$ & $(4.17\pm0.13)\times10^{16}$ & $3.6\times10^{-5}$ \\
OCS         & $-17.9\pm0.1$ & $2.5\pm0.2$ & $(2.09\pm0.15)\times10^{13}$ & $2.8\times10^{-8}$ & $(1.51\pm0.11)\times10^{13}$ & $1.3\times10^{-8}$ \\
O$^{13}$CS  & $-18.4\pm0.1$ & $1.5^b$      & $(8.91\pm1.02)\times10^{12}$ & $1.2\times10^{-8}$ & $(6.17\pm0.81)\times10^{12}$ & $5.4\times10^{-9}$ \\
SO          & $-18.0\pm0.1$ & $2.4\pm0.1$ & $(5.62\pm0.07)\times10^{13}$ & $7.4\times10^{-8}$ & $(8.32\pm0.11)\times10^{13}$ & $7.2\times10^{-8}$ \\
SO$_2$      & $-18.4\pm0.2$ & $3.5\pm0.5$ & $(8.51\pm1.17)\times10^{15}$ & $1.1\times10^{-5}$ & $(7.24\pm0.98)\times10^{14}$ & $6.3\times10^{-7}$ \\
$^{33}$SO   & $-18.4\pm0.1$ & $2.1\pm0.3$ & $(3.80\pm0.30)\times10^{12}$ & $5.0\times10^{-9}$ & $(5.89\pm0.46)\times10^{12}$ & $5.1\times10^{-9}$ \\
HNCO        & $-17.6\pm0.2$ & $2.3\pm0.5$ & $(3.63\pm0.55)\times10^{12}$ & $4.8\times10^{-9}$ & $(3.72\pm0.58)\times10^{12}$ & $3.2\times10^{-9}$ \\
H$_2$CO     & $-18.3\pm0.1$ & $3.0\pm0.1$ & $(1.15\pm0.11)\times10^{13}$ & $1.5\times10^{-8}$ & -- $^c$                       & -- \\
$^{13}$CS   & $-18.2\pm0.3$ & $2.7\pm0.8$ & $(5.89\pm1.41)\times10^{11}$ & $7.8\times10^{-10}$ & $(8.32\pm2.00)\times10^{11}$ & $7.2\times10^{-10}$ \\
\hline
\end{tabular}
%\begin{flushleft}
\footnotesize
$^a$ Standard reference value, as no direct measurements of the H$_2$ column density are available.
$^b$ Fixed value, for fitting convergence.
$^c$ Fitting not converging.
%\end{flushleft}
\end{table*}

The fit indicates that most species are optically thin, with $\tau$ values ranging from 0.01 to 0.03. The exceptions are CO ($\tau=0.56$) and $^{13}$CO  ($\tau=0.22$), which exhibit slightly flat-topped line profiles consistent with moderately thick emission.  While these two species present the highest column densities, and \texttt{MADCUBA} provides opacity-corrected values, these results should be treated as conservative lower limits, as optical thickness, derived from a single transition, is likely underestimated. Despite these cautions, these values yield  a [CO/$^{13}$CO] of $\sim$3--4, a remarkably low value that suggests a considerable degree of $^{13}$C enrichment from CNO-processed material.

To derive fractional abundances, we referred all the column density values to that of CO, adopting a standard $X$(CO/H$_2$) = 10$^{-4}$. This value is consistent with the typical range observed in cooler super- and hypergiant stars ($0.4-5.3\times10^{-4}$, \citealt{sin22}).  Interestingly, SO$_2$ is the third most abundant species after CO and $^{13}$CO, regardless of the assumed temperature. At $T_\mathrm{ex}$=50 K, $X$(SO$_2$)=$1.1\times10^{-5}$ is  close to the cosmic elemental abundance of sulfur (S/H $\sim 1.5\times10^{-5}$). This indicates that SO$_2$ is the primary sulfur carrier in this source. In other words, most of the available elemental sulfur is locked into SO$_2$, implying no significant depletion of sulfur into dust grains. Large amounts of sulfur returned to gas-phase have also been observed in some AGB envelopes \citep{san97}. 

Values of the molecular ratio SO$_{2}$/SO $\sim 0.1$ are found in cold cores \cite[Barnard 1b,][]{Fuente2016} and photodissociation regions \cite[Horsehead PDR,][]{Riviere-Marichalar2019}, while higher ratios SO$_{2}$/SO $\sim 1$ are measured in the bowshocks associated to bipolar outflows \citep{Bachiller1997} and dust traps in protoplanetary disks \cite[Oph IRS 48,][]{Booth2021}. Likewise, the oxygen-rich CSEs of cool supergiants, such as VY CMa \citep{ziu07,ziu09}, NML Cyg \citep{sin22} and IRC+10420 \citep{qui16}, typically present SO more abundant than, or comparable to, SO$_2$ (as predicted by standard chemical models, \citealt{wil87}). On the other hand, SO$_{2}$/SO abundance ratios well above 1 have only been observed in hot cores \cite[see, e.g.,][]{Jimenez-Serra2012, Esplugues2013, Fuente2021} and locally in the outflows of some evolved stars \citep{cla00,ada13}. Therefore, HD~87643 stands out as an extreme case, with SO$_2$  found to be 10--150 times more abundant than SO. The presence of shocks, frequently observed in other massive stars \citep{riz14} may explain this out-of-equilibrium chemistry. This scenario is further supported by the detection of HNCO, a common tracer of low-velocity shocks. This molecule can be produced via  sputtering of dust grains  \citep{rod10, yu18} and has been reported in the CSEs of some evolved stars \citep{vel15}.

\subsection{Sulfur chemistry in HD~87643}

The observed abundances are consistent with the emission in HD 87643 coming from a warm and dense layer of circumbinary material probed by dust continuum emission. Given the spatial resolution of our observations, it is not possible to fully constrain the size and physical properties of the emitting region. Instead, we performed a chemical modeling of the region to find the physical parameters that best describe it. We used the gas-grain chemical code \texttt{NAUTILUS}\footnote{Version 2.0.0 with its corresponding updated chemical network \citep{Wakelam2024}.} \citep{Ruaud2016} to reproduce the fractional abundances of OCS, SO, SO$_{2}$, HNCO, and H$_{2}$CO. \texttt{NAUTILUS} takes into account the chemical processes and interactions that occur in the gas-phase, the surface of icy grains, and ice mantles, following the rate equation approach to compute the chemical composition through time given an initial set of physical and chemical conditions.

\begin{table}[h!]
\centering
\caption{List of initial abundances with respect to atomic hydrogen for the chemical modeling of HD 87643.}
\label{tab:initialAbs}
\begin{tabular}{lc}
\hline
\hline
Species & Initial abundance (/H) \\
\hline
He      & $9.0\times 10^{-2}$ \\
N       & $6.2\times 10^{-5}$ \\
O       & $4.3\times 10^{-4}$ \\
H$_{2}$ & $5.0\times 10^{-1}$ \\
C$^{+}$ & $1.7\times 10^{-4}$ \\
S$^{+}$ & $1.5\times 10^{-5}$ \\
Si$^{+}$& $8.0\times 10^{-9}$ \\
Fe$^{+}$& $3.0\times 10^{-9}$ \\
Na$^{+}$& $2.0\times 10^{-9}$ \\
Mg$^{+}$& $7.0\times 10^{-9}$ \\
P$^{+}$ & $2.0\times 10^{-10}$ \\
Cl$^{+}$& $1.0\times 10^{-9}$ \\
F       & $6.7\times 10^{-9}$ \\
\hline
\end{tabular}
\end{table}

\begin{figure}
    \centering
    \includegraphics[width=\columnwidth]{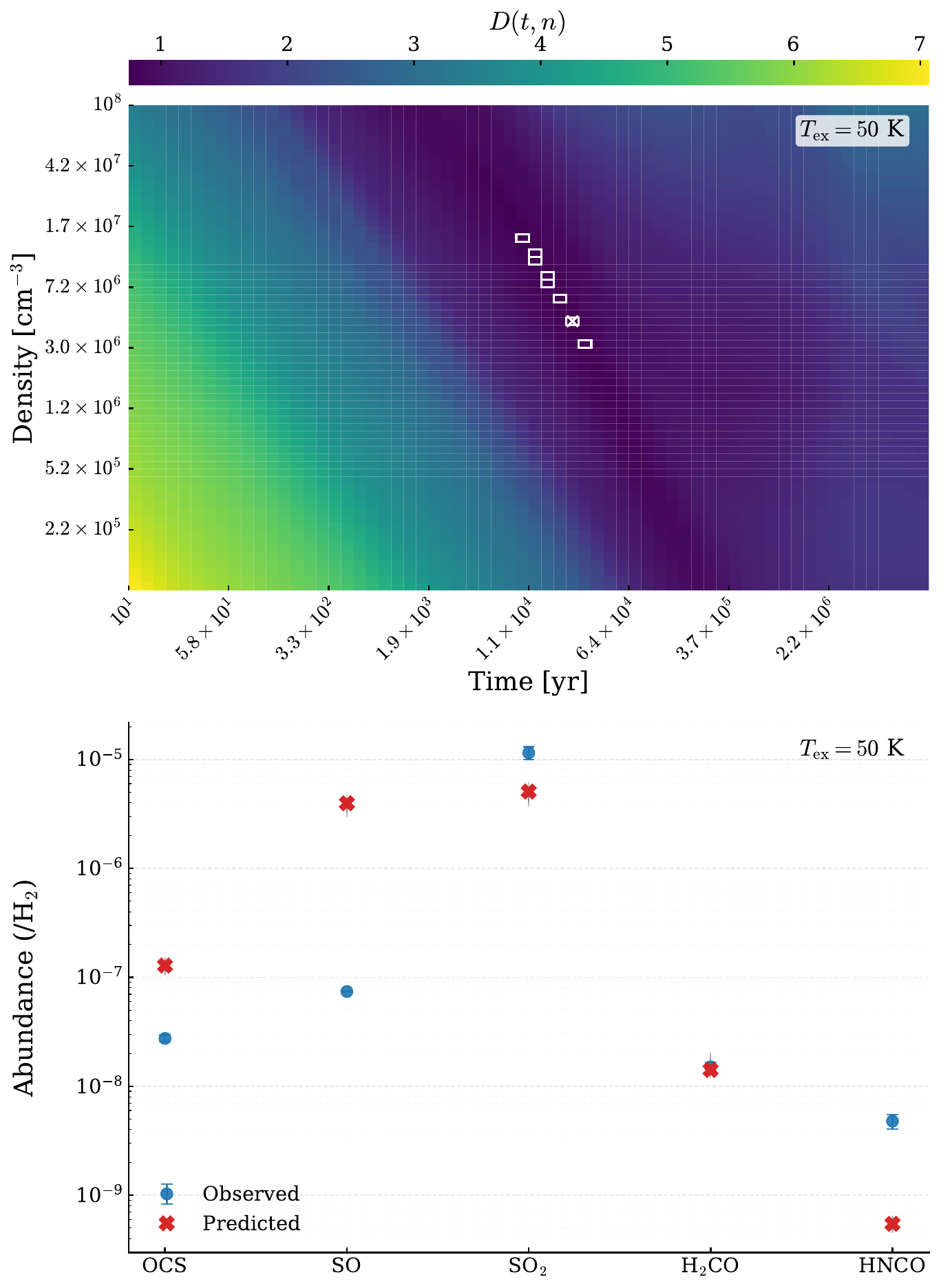}
    \caption{\emph{Top:} grid of models with varying gas densities and chemical time, and their corresponding distances of disagreement (Eq.\ref{eq:distDis}), for $T$=50 K. Best fitting models are marked with white squares. \emph{Bottom:} comparison between observed abundances (blue dots) and predictions (red crosses) from the best model (indicated with a white cross in the top panel).}
    \label{fig:modelResults}
\end{figure}

For the model we assumed that the molecular emission comes from a gas that fills the beam with $T_{\rm k}=50$ K. This temperature is consistent with the $T_{\rm ex}$ derived from the H$_{2}$CO and OCS lines and the dust temperature ranges of the cool dust in the SED of HD 87643 (see Fig. \ref{fig:fig3}). To find the H$_2$ number density and chemical time that best fit the observed abundances, we ran a grid of models with varying densities in the range $10^{5}-10^{8}$ cm$^{-3}$. This range encompasses uncertainties in the size of the emitting region and in the dust mass assuming a spherical geometry. The initial chemical abundances set in the grid are listed in Table \ref{tab:initialAbs}. The assumed ratio C/O $\sim 0.4$, characteristic of an O-rich environment, was set according to the spectroscopic results towards the sample of B-type supergiants reported in \citet{Wessmayer2022}. Given the high abundance of SO$_{2}$ at $T_{\rm ex}=50$ K, close to the cosmic sulfur abundance, the grid of chemical models includes undepleted sulfur $[{\rm S}] = 1.5\times 10^{-5}$. Finally, we adopted a standard gas-to-dust ratio of 100. To assess the quality of a model, we used the distance of disagreement $D$ \citep{Wakelam2024}:

\begin{equation}\label{eq:distDis}
    D(t,n) = \frac{1}{N}\sum^{N}_{i}|\log(X_{\rm mod,i}(t,n))-\log(X_{\rm obs,i})|
\end{equation}

where $t$ and $n$ are the chemical time and gas density, $N$ is the number of species, and $X_{\rm mod,i}$ and $X_{\rm obs,i}$ are the predicted and observed abundances of the $i$-th species (OCS, SO, SO$_{2}$, HNCO, and H$_{2}$CO; derived at $T_{\rm ex}=50$~K, Table \ref{tab:detections}). The distance of disagreement is shown in the top panel of Fig. \ref{fig:modelResults}. The $(t,n)$ pairs in the lowest 0.2 percentile of $D(t,n)$ represent the best-fit models (white markers, with the minimum marked by a cross).

We found that the emitting region is best characterized by a density range $(0.3-1.6)\times 10^{7}$ cm$^{-3}$ and a chemical time $(0.9-2.7)\times 10^{4}$ yr. This timescale represents the time required for the chemical network to evolve from the initial conditions to the stage that best reproduces the observed molecular abundances, and sets a lower boundary on the age of the external H$\alpha$ nebula---which is necessarily older than the circumbinary structure. The corresponding predicted abundances are shown on the bottom panel of Fig. \ref{fig:modelResults}. We also attempted to reproduce the abundances with a standard C/O$\sim$0.6 (typical of AGB stars), and found little difference: slightly higher times and lower densities, still within the previous ranges and with an overall worse $D(t,n)$.  
  
Although the abundance of all species could not be fitted simultaneously, we obtained a reasonably good agreement for the OCS, H$_{2}$CO, HNCO, and SO$_{2}$ species, with predictions within an order of magnitude from the observed values. This model therefore accounts for the high abundance of sulfur detected in SO$_{2}$ toward HD 87643. According to the chemical network, the radical-neutral gas-phase reaction ${\rm OH} + {\rm SO}\rightarrow {\rm H} + {\rm SO_{2}}$ is the main mechanism that produces this molecule, whereas it is destroyed by the ion-neutral reaction $ {\rm SO_{2}} + {\rm {H_{3}}^{+}} \rightarrow {\rm H_{2}} + {\rm HS{O_{2}}}^{+}$. Our results suggest that the observed molecular emission comes from a relatively dense and cool structure surrounding HD 87643. Furthermore, the short chemical timescale derived from the best-fitting model is compatible with a transient, rapidly evolving environment, linking the gas to a recent mass-ejection or a non-conservative binary mass transfer episode. Higher-resolution observations will be needed to narrow down the exact origin of this structure.

\section{Discussion}\label{sec4}

The proposed chemical model represents an initial step toward characterizing the molecular chemistry of HD 87643. It successfully reproduces the observed abundance of SO$_2$, thereby accounting for most of the gas-phase sulfur. Notably, the predicted SO$_2$ abundance is reached on timescales of $\sim$10$^4$ yr, which is entirely consistent with the post-MS nature of the supergiant phase \citep{gro14}. However, the model overestimates the abundance of SO by a large margin. This discrepancy suggests that alternative formation and destruction pathways need to be considered, likely influenced by the UV-driven photochemistry expected in the outskirts of a B3I-type star. 

In this respect, the isotopic composition of the gas poses another chemical puzzle. The analysis of our observations (Table \ref{tab:lte}) reveals an exceptionally low $^{32}$SO/$^{33}$SO ratio of $\sim15$. Sulfur has four stable isotopes, $^{32}$S, $^{34}$S, $^{33}$S and $^{36}$S, with abundance ratios of 95.02 : 4.21 : 0.75 : 0.021 in the Solar System \citep{and89}. Therefore, if the observed $^{32}$SO/$^{33}$SO ratio reflects the underlying elemental abundances,  it implies a $^{33}$S enrichment of nearly an order of magnitude relative to the Solar $^{32}$S/$^{33}$S of $\sim$127. Such a value would be anomalous compared to other Galactic environments, which are broadly consistent with Solar values, ranging from 70$\pm$16 in the Central Molecular Zone, 88$\pm$21 in the local ISM, and 105$\pm$19 in the outer Galaxy \citep{yan23}. Similarly, among evolved stars, carbon-rich AGB stars like IRC+10216 show compatible ratios (121$\pm$15, \citealt{mau04}). Obviously, opacity would help explain the observed ratio if the $^{32}$SO line were heavily saturated, requiring $\tau\sim10$. This would also contribute to partially reconciling the observations with the \texttt{NAUTILUS} results. However, this seems not to be the case, as the line profile appears narrow and remarkably Gaussian, showing no hints of high optical depth. 

Therefore, alternative mechanisms must be explored to explain this anomalous ratio. One such mechanism is nucleosynthetic enrichment of $^{33}$S. A notable case is the oxygen-rich AGB star R Dor, with a $^{32}$S/$^{33}$S ratio of 68$\pm$22 \citep{dan20}, well below the Galactic trend and attributed to enrichment during the AGB phase via the weak $s$-process \citep{and89,wal24}.  However, this process alone is unlikely to fully account for such a low SO/$^{33}$SO ratio in {HD~87643}. The weak $s$-process in massive stars, taking place during  core helium- and shell carbon-burning  \citep{cou74, lan89, rai91, pig10} , is not particularly efficient at accumulating $^{33}$S  relative to other isotopes, as the neutron capture flow is recycled to lighter elements via the reaction $^{33}$S(n,$\alpha$)$^{30}$Si, leading to a modest $^{33}$S overproduction factor \citep{rei00}. In consequence, evolutionary models of 13--120 M$_\odot$ stars by \cite{lim18} do not predict a significant $^{33}$S enrichment with respect to $^{32}$S. Similarly, nucleosynthetic models of classical novae predict moderate $^{33}$S enrichment in the ejecta \citep{jos01}, but still inssuficient to explain values as low as the one observed in HD 87643.

A simpler, albeit more speculative possibility not involving nucleosynthesis is that the anomalous molecular abundance ratio arises from mass-independent fractionation (MIF), rather than an intrinsically low sulfur isotopic ratio. In {HD~87643} SO$_2$ is the main sulfur reservoir. Conversely, its isotopologue $^{33}$SO$_2$ is not detected, despite the frequency setup covering several of its transitions. Using \texttt{MADCUBA} we determine a $3\sigma$ upper limit  $N(^{33}\mathrm{SO}_2) < 6\times10^{13}$ cm$^{-2}$. This upper limit implies a close-to-standard (Solar) $^{32}$S/$^{33}$S ratio ($>$140). 

In this scenario,  $^{32}$SO and $^{33}$SO would be daughter species---as opposed to classical models of AGB winds, \citealt{che06}---,mainly produced via the photolysis of their corresponding parent molecules, $^{32}$SO$_2$ and $^{33}$SO$_2$,  by UV photons in the wavelength range 190--220 nm \citep{whi13}:

\begin{equation}
\begin{split}
    {}^{32}\mathrm{SO}_2 + h\nu &\rightarrow {}^{32}\mathrm{SO} + \mathrm{O} \\
    {}^{33}\mathrm{SO}_2 + h\nu &\rightarrow {}^{33}\mathrm{SO} + \mathrm{O}
\end{split}
\end{equation}

Given its high abundance  ($\sim$10$^{-5}$ at $T_\mathrm{ex}$=50 K),  $^{32}$SO$_2$, the dominant isotopologue, would likely be self-shielded against the intense UV radiation from the central B[e] star, effectively limiting photodissociation to the directly exposed gas layers. The much less abundant $^{33}$SO$_2$, by contrast, would remain optically thin and, with its slightly different absorption cross section \citep{dan08}, would get photodissociated more efficiently through the entire cloud volume. Consequently, the relative production rate of $^{33}$SO would significantly exceed that of $^{32}$SO, explaining the anomalous abundance ratio without invoking exotic nucleosynthetic scenarios. Interestingly, MIF driven by SO$_2$ photolysis has been widely discussed in the context of the Archaean Earth's atmosphere and meteoritic sulfur anomalies \citep{far00, far01, mas11}. Laboratory studies explicitly attribute observed $^{33}$S excesses to isotope-selective self-shielding, which yields photodissociation products enriched in heavier isotopes \citep{lyo07,dan08,end15}. Therefore, the circumstellar gas of {HD~87643} may represent an extreme case of sulfur MIF, serving as a unique testbed for isotope-selective photochemistry.

\section{Conclusions}

The study presented here reports the first detection of sulfur oxides and other sulfur-bearing species in an evolved early-type massive star, significantly expanding the molecular inventory of B[e] supergiants. Whether HD 87643 represents a rare chemical anomaly---enabled by its unique nature and complex environment---or is instead an exemplar of a broader phenomenology remains to be determined. In either case, the observed abundances and isotopic ratios point to a highly transient chemistry, far from steady-state chemical equilibrium, governed by the continuous interplay between mass-loss, UV radiation, and potentially shocks.

By combining radiative transfer modelling, SED reconstruction, and chemical simulations, our analysis has tackled HD 87643 in a multifaceted manner that is readily applicable to similar sources. Future high-resolution, multi-transition ALMA observations will be crucial to constrain the spatial distribution of the gas, trace back its origin, and complete the picture of sulfur chemistry. More broadly, understanding how molecules form and survive near evolved supergiants provides an valuable proxy for the chemical enrichment of the early Universe, where massive stars played a crucial role in spreading the chemical elements needed for the formation of planetary systems and (ultimately) the emergence of life.

%% Please use the acknowledgment and contribution environments. This will 
%% be anonomyized when the "anonymous" style option is used. 
\begin{acknowledgments}
This paper makes use of the following ALMA data: ADS/JAO.ALMA\#2023.1.01688.S ALMA is a partnership of ESO (representing its member states), NSF (USA) and NINS (Japan), together with NRC (Canada), NSTC and ASIAA (Taiwan), and KASI (Republic of Korea), in cooperation with the Republic of Chile. The Joint ALMA Observatory is operated by ESO, AUI/NRAO and NAOJ. CB acknowledges financial support from INAF – Ricerca Fondamentale 2024 Mini Grant program (Ob. Fu. 1.05.24.07.02), from grant CEX2021-001131-S funded by MICIU/AEI/ 10.13039/501100011033 and from grant INFRA24023 (CSIC4SKA) funded by CSIC. AF thanks project PID2022-137980NB-I00 funded by the Spanish
Ministry of Science and Innovation/State Agency of Research MCIN/AEI/10.13039/501100011033 and by “ERDF A way of making Europe”. This work is co-funded by ERC grant SUL4LIFE, GA No. 101096293 Funded by the European Union. Views and opinions expressed are however those of the author(s) only and do not necessarily reflect those of the European Union or the European Research Council Executive Agency. Neither the European Union nor the granting authority can be held responsible for them. This work was supported by PID2022137779OB-C41 funded by MCIN/AEI/10.13039/501100011033 by “ERDF A way of making Europe”. This research has made use of the Spanish Virtual Observatory (\url{https://svo.cab.inta-csic.es}) project funded by
MCIU/AEI/10.13039/501100011033/ through grant PID2023-146210NB-I00.
\end{acknowledgments}

%\facilities{ALMA(STIS), Swift(XRT and UVOT), AAVSO, CTIO:1.3m, CTIO:1.5m, CXO}

%\software{astropy \citep{2013A&A...558A..33A,2018AJ....156..123A,2022ApJ...935..167A},  
%          Cloudy \citep{2013RMxAA..49..137F}, 
%          Source Extractor \citep{1996A&AS..117..393B}
%          }


\appendix

\section{Spectral Energy Distribution building}
\label{methods:sed}

We constructed the spectral energy distribution (SED) of HD 87643 from optical to radio wavelengths to constrain the physical properties of its circumstellar environment, and specifically the circumbinary dust temperature. The SED incorporates photometry compiled from multiple surveys and catalogues (Table \ref{tab:photometry}), our ALMA measurement of the $1$ mm flux density ($S=216\pm2$ mJy), and a detection at 943 MHz from the ASKAP Evolutionary Map of the Universe survey (tile EMU$\_$0954-55, observed in SB51428, \citealt{hop25}), for which we determined a flux density $S=0.006\pm0.004$ mJy using standard aperture photometry.

\begin{table}[h!]
\centering
\caption{Photometric data for HD~87643.}
\label{tab:photometry}
\begin{tabular}{l l l}
\hline
\textbf{Survey} & \textbf{Catalogue} & \textbf{Filters} \\
\hline
Gaia    & I/355/gaiadr3     & $G_{BP}$, $G$, $G_{RP}$ \\
2MASS   & II/246/out        & $J$, $H$, $K_s$ \\
AKARI   & II/297/irc        & AKARI:S9W, AKARI:L18W \\
IRAS    & II/125/main       &  IRAS:12, IRAS:25, IRAS:60, IRAS:100   \\
MSX     & V/114/msx6\_gp     & MSX:B1, MSX:B2, MSX:A, MXS:C, MSX:D, MXS:E \\
ISO     & ISO/legacy        & ISO:180 \\
ALMA    & This work         & Band 6 ($230$ GHz) \\
ASKAP   & EMU Survey        & $943$ MHz \\
\hline
\end{tabular}
\end{table}

We modeled the SED following the methodology described in \cite{riz25}, representing the system as a combination of analytical components with physically-motivated constraints. We kept the model intentionally simple, as the goal is to obtain a first-order estimate of the circumbinary dust temperature. In the case of HD~87643:

\begin{itemize}
    \item We model the primary star as an ideal black body with a temperature constrained to the 16,000--20,000 K range, consistent with the B3 spectral type of {HD~87643} \citep{sho90}. The choice of a black body over a more complex synthetic model has negligible impact on the derived dust temperatures.
    \item We model the hot circumprimary disk as a series of concentric, modified blackbodies with a temperature gradient. We constrain its innermost temperature to 1,500–2,500 K, consistent with previous estimates (\citealt{mil09}) and the detection of CO overtone emission (\citealt{mar18}), and vary its emissivity index $\beta$ from 0 (inner edge) to 0.7 (outer edge).
    \item We model the larger, cooler envelope as a dust component with an inner temperature of 100–300 K, consistent with \cite{mil09}, and a fixed emissivity index of $\beta=2$, indicative of small grains.
    \item Finally, we add an idealized stellar wind to account for the cm-wave flux density, following the prescription of \cite{pan75}. We assume a mass-loss rate $\dot M=1.7\times10^{-5}$ M$_\odot$ yr$^{-1}$, a terminal wind velocity $v_\infty=1000$ km s$^{-1}$, and a plasma temperature of $T=10^4$ K, able to reproduce the observed flux density at 900 MHz.
\end{itemize}

We did not model the secondary companion and its putative unresolved envelope reported by \cite{mil09}, as their contribution to the total flux is expected to be negligible. Prior to the fitting, the observed SED was dereddened considering a combination of interstellar and circumstellar extinction, taking the maximum possible absorption compatible with the spectral slope of a B3 star, an assumption justified by the large amounts of dust enshrouding the object. The adopted $A_V$ = 3.8 is compatible with literature estimates of interstellar extinction (\citealt{oud98}).

\begin{table}[h]
\centering
\caption{Best-fitting SED parameters with uncertainties.}
\label{tab:parameters}
\begin{tabular}{lll}
\toprule
\textbf{Parameter} & \textbf{Value} & \textbf{Unit} \\
\midrule
\multicolumn{3}{l}{\textit{Stellar Photosphere}} \\
$T_{\text{eff}}$ & $18500 \pm 1200$ & K \\
$R_{*}$ & $20 \pm 2$ & $R_{\odot}$ \\
\midrule
\multicolumn{3}{l}{\textit{Hot Disk}} \\
$T_{\text{inner}}$ & $2000 \pm 300$ & K \\
$\beta_{\min}$ & 0.01 & --- \\
$\beta_{\max}$ & 0.7 & --- \\
\midrule
\multicolumn{3}{l}{\textit{Cool Disk}} \\
$T_{\text{inner}}$ & $80 \pm 10$ & K \\
$\beta$ & 2.0 & --- \\
\midrule
\multicolumn{3}{l}{\textit{Stellar wind}} \\
$\dot{M}$ & $(1.7\pm1.0) \times 10^{-5}$ & $M_{\odot}\,\text{yr}^{-1}$ \\
$v_{\exp}$ & 1000 & km\,s$^{-1}$ \\
\bottomrule
\end{tabular}
\end{table}

We performed a $\chi^2$ minimization to find the best-fit model parameters, following the method described in \cite{riz25}. The resulting parameters and associated uncertainties are reported in Table \ref{tab:parameters}. Despite its simplicity, the model reproduces the observed emission through the electromagnetic spectrum reasonably well (Fig. \ref{fig:fig3}), and yields several key insights. First, we constrain the circumbinary envelope temperature to 20--80 K, a range compatible with the numerous molecular species detected toward the star. Furthermore, the model shows that the stellar wind's contribution at millimeter wavelengths is almost negligible. Consequently, we can attribute $\sim$95\% of the observed ACA flux density to thermal dust emission, effectively allowing for an estimate of the total dust mass in the envelope, of $(3.7-7.3)\times$10$^{-2}$ M$_\odot$ (for $T_d$=50 K, $\kappa_0$=1 and 0.5 g cm$^{-2}$, respectively--$\kappa_0$ being the reference dust mass opacity).

\section{Distance and evolutionary status of HD~87643}
\label{methods:evol}

While generally accepted to be an evolved supergiant, the evolutionary status of  {HD~87643} has long been controversial. Two main methods  can discriminate between the pre- and post-Main Sequence scenarios for B[e] stars: luminosity considerations and the analysis of isotopic ratios \citep{kra09}. Luminosity calculations were previously hindered due to the lack of reliable distance measurements, with a wide range of estimates available in the literature (see \citealt{mil09}, section 3.1, for a discussion).  However, the Gaia DR3 parallax of $\epsilon=$0.6293$\pm$0.1227 mas yields a distance of 1.58$^{+0.39}_{-0.25}$ kpc, settling the supergiant classification: our SED model results ($R_*$=20 R$_\odot$, $T_\mathrm{eff}=18500$ K) yield a luminosity of $\log(L/L_\odot)$=4.6, placing the source on the low-luminosity end of the  B[e] supergiant region of the HR diagram \citep{zic06}.

Chemical analysis provides a second, independent assessment. The evolution of massive stars naturally produces $^{13}$C as a byproduct of the CNO cycle. Therefore, the [$^{12}$CO/\ensuremath{^{13}\mathrm{CO}}] $\approx$ [$^{12}$C/$^{13}$C] ratio is generally a powerful probe of evolutionary status.  From the column densities in Table \ref{tab:lte}, we derive an isotopic ratio [$^{12}$CO/\ensuremath{^{13}\mathrm{CO}}]$\sim 3$. This is an exceptionally low value, far below the standard ISM value of $\sim$70 \citep {wil94}. However, this result must be treated with caution. Both the $^{12}$CO and \ensuremath{^{13}\mathrm{CO}}\,  lines show slightly flat-topped profiles, suggesting they are at least moderately optically thick. Although the \texttt{MADCUBA} fit corrects for opacity, its effect is very likely underestimated. Consequently, the derived isotopic ratio of $\sim3$ must be regarded as an strict lower limit. However, even if the real ratio is off by a factor of five, it would remain fully compatible with a post-Main Sequence status, as B[e] supergiants typically show [$^{12}$CO/\ensuremath{^{13}\mathrm{CO}}] values in the range 4--20 \citep{kra19}. In any case, to properly constrain the optical depths and determine the true isotopic ratio, multi-transition observations are required.

The evolved scenario is further supported by the intriguing resemblance between {HD~87643} and known Galactic LBVs. Early morphological analyses of its circumstellar nebula first suggested this link \citep{sur81, sur83}, and subsequent photometric monitoring has revealed significant variability: a long-term decline of $\sim0.9$ mag over two decades, superposed with shorter fluctuations of $~$0.5 mag on timescales of a few months \citep{poj09,mil09}. While these features do not strictly mirror the S Doradus cycles characteristic of bona fide LBV stars, {HD~87643} may represent a rare hybrid case consistent with the intricate evolutionary overlap between LBVs and sgB[e] stars.

\bibliography{manuscript}{}
\bibliographystyle{aasjournalv7}



\end{document}